\documentclass[angeo]{copernicus}

\begin{document}

\newcommand\ion[2]{\textup{{#1}}~\textsc{{#2}}}%
\def\aap{A\&A}%
\def\aaps{A\&AS}%
\def\apj{ApJ}%
\def\solphys{Sol.~Phys.}%

\title{Diagnostics of active and eruptive prominences through hydrogen and helium lines modelling}

\author[1,*]{N. Labrosse}
\author[2]{J.-C. Vial}
\author[2]{P. Gouttebroze}

\affil[1]{Institute of Mathematical and Physical Sciences, University of Wales Aberystwyth, UK}
\affil[2]{Institut d'Astrophysique Spatiale, CNRS -- Universit\'e Paris XI, Orsay, FR}
\affil[*]{now at: Department of Physics and Astronomy, University of Glasgow, UK}

\runningtitle{Diagnostics of active and eruptive prominences}

\runningauthor{N. Labrosse et al.}

\correspondence{N. Labrosse\\ (nicolas@astro.gla.ac.uk)}

\received{}
\pubdiscuss{} 
\revised{}
\accepted{}
\published{}


\firstpage{1}

\maketitle

\begin{abstract}
In this study we show how hydrogen and helium lines modelling can be used to make a diagnostic of active and eruptive prominences.
One motivation for this work is to identify the physical conditions during prominence activation and eruption. 
Hydrogen and helium lines are key in probing different parts of the prominence structure and inferring the plasma parameters.
However, the interpretation of observations, being either spectroscopic or obtained with imaging, is not straightforward. Their resonance lines are optically thick, and the prominence plasma is out of local thermodynamic equilibrium due to the strong incident radiation coming from the solar disk.
In view of the shift of the incident radiation occurring when the prominence
plasma flows radially, it is essential to take into account velocity fields in the
prominence diagnostic.
Therefore we need to investigate the effects of the radial motion of the prominence plasma on hydrogen and helium lines. 
The method that we use is the resolution of the radiative transfer problem in the hydrogen and helium lines out of local thermodynamic equilibrium.
We study  the variation of the computed integrated intensities in H and He lines with the radial velocity of the prominence plasma.
We can confirm that there exist suitable lines which can be used to make a diagnostic of the plasma in active and eruptive prominences in the presence of velocity fields.
\end{abstract}

\introduction
If we want to understand the energy budget and the physical mechanisms acting in erupting prominences, it is crucial to measure the full velocity vector.
The full velocity vector may be inferred but requires at least the radial velocity. 
There are various techniques available to measure velocities in prominences: imaging measurements, leading to the apparent motion of the structure in the plane of the sky; Doppler shifts, yielding the velocity along the line of sight (LOS); and Doppler dimming and brightening. The latter varies with the radial velocity of the emitting plasma. 
We will show that by studying the Doppler dimming or brightening effect on the hydrogen and helium lines, it is possible to make a consistent diagnostic of the prominence plasma, taking into account the radial velocity.

What are the interesting factors and challenges associated with the H and He lines in this type of study? One interesting characteristic of these lines is that they form in different parts of the prominence.
H Lyman lines, from the lower to the higher members of the line series, can be used to access several regions of the structure, using either their optically thick core (which reveals the prominence fine structure close to the prominence--corona boundary) or their optically thin wings \citep[e.g.,][]{2007A&A...472..929G,2007ASPC..368..271H}. Note, however, that the optically thin part of the line wings implies that several different elements of the prominence structure are integrated along the LOS.
The same reasoning also applies to the helium resonance lines such as \ion{He}{i} $\lambda\lambda$ 584~\AA\ and 537~\AA, and \ion{He}{ii} $\lambda\lambda$ 304~\AA\ and 256~\AA\ \citep{2001A&A...380..323L}. 
All these EUV resonance lines are observed by imagers and spectrometers on board SOHO, CORONAS-F/SPIRIT, Hinode, STEREO, and in the future, on SDO.
The challenge is that these resonance lines are optically thick and that the prominence plasma is out of LTE (Local Thermodynamic Equilibrium), and therefore the plasma diagnostic is complex.
In short, non-LTE radiative transfer calculations including velocity fields are needed to build realistic prominence models.

In the following we briefly describe the non-LTE radiative transfer calculations performed to predict the total radiation emitted by hydrogen and helium in active and eruptive prominences. We then present our results for a selection of lines: \ion{He}{ii} $\lambda$304~\AA, \ion{He}{i} $\lambda$584~\AA, and  hydrogen H$\alpha$.

\section{Effect of radial motions on spectral line profiles}
Let us give  a simple description of the effect of radial motions on the emitted spectrum. Considering a simple 2-level atom whose lower level is excited by the radiation coming from the Sun, the absorption profile of the radiative transition between the two levels gets out of resonance with the incident radiation when this atom is moving outwards, due to the Doppler effect. We namely have a Doppler dimming effect when the incident line is in emission, or a Doppler brightening effect if the incident line is in absorption.

In a more realistic situation, an atom is likely to have more than two energy levels. Consequently, coupling effects take place between the atomic levels. For that reason, a combination of Doppler dimming and brightening can occur, just as it happens when the coupling between the first two excited levels of hydrogen is taken into account \citep{1987SoPh..110..171H,1997SoPh..172..189G}.

This is the same Doppler effect which is also used for the diagnostic of the radial component of the solar wind \citep[see, e.g.,][]{1982ApJ...256..263K}.

The main factors determining the effects of the radial motions on the emitted prominence spectrum are the line formation mechanisms, including the relative contributions of collisional and radiative excitation, and the characteristics of the incident radiation (whether it is in absorption or in emission, the strength of the line, etc). More discussion can be found in \citet{1987SoPh..110..171H,1997A&A...325..803G,2006IAUJD...3E..47L,2007A&A...463.1171L,2007ASPC..368..337L}.

\section{Calculations}
We now summarize the main features of the method that we use to compute the spectrum emitted by H and He in prominences.
More details are given in \citet{2007A&A...463.1171L}.

\begin{figure}[t]
\vspace*{2mm}
\begin{center}
\includegraphics[width=8.3cm]{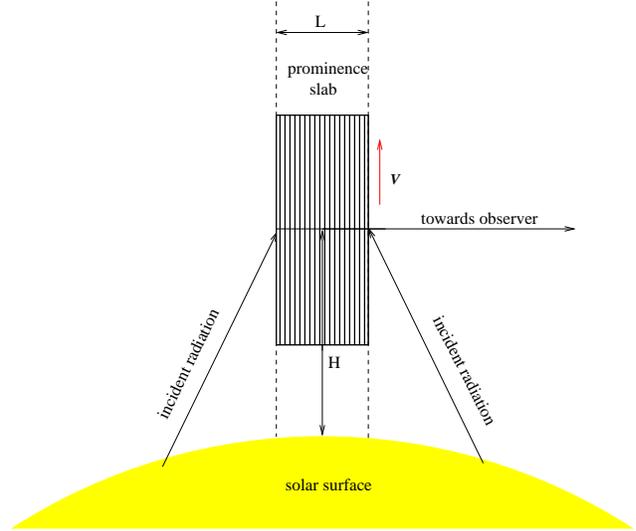}
\end{center}
\caption{Schematic representation of the 1D plane-parallel prominence slab of geometrical width $L$, altitude $H$ above the solar surface, moving radially at a velocity $V$.}
\label{prom:fig}
\end{figure}

The prominence itself is represented as a 1D plane-parallel slab standing vertically above the solar surface (see Fig.~1). We fix the temperature, the gas pressure, the column mass, the altitude, and the radial velocity. Then we solve the coupled statistical equilibrium and radiative transfer equations, to obtain the energy level populations and the emitted radiation.

The temperature and the gas pressure are determined from two types of models. The first type yields isothermal and isobaric prominences, where the temperature and the pressure are constant throughout the whole prominence slab. 
This is similar to what has been used so far by the various authors who studied the Doppler effect in solar prominences.
The second type of models includes a prominence-to-corona transition region, with a temperature and pressure gradient. The pressure and temperature profiles are taken from the work of \citet{1999A&A...349..974A}.
The pressure profile is given by:
\begin{equation} 
  p(m) = 4 p_c \frac{m}{M} \left( 1-\frac{m}{M} \right) + p_0 \ , \label{varp}
\end{equation}
where $M$ is the total column mass of the prominence, $m$ is the column
mass, $p_0$ is the coronal pressure at the outer boundary, and $p_c =
p_{cen} - p_0$, with $p_{cen}$ the central gas pressure. Note that in equation~(\ref{varp}), $p_c > 0$, so the central pressure is always greater than the coronal pressure.
The temperature profile is given by:
\begin{equation}
  T(m) = T_{cen} + \left( T_{tr}-T_{cen} \right) \left[ 1-4\frac{m}{M} \left(1-\frac{m}{M} \right) \right] ^\gamma \ ,\label{tempprof}
\end{equation}
where $T_{cen}$ and $T_{tr}$ are the temperatures at the centre of the slab and at the outer boundary respectively.
This temperature profile is semi-empirical and does not result from a physical derivation.
The $\gamma$ exponent in Eq.~(\ref{tempprof}) is a free parameter and it tells us how steep the temperature gradient is. In other words, the higher $\gamma$, the steeper the gradient, and the less extended the transition region is. 

We have computed 4 different types of prominence models, either a isothermal isobaric prominence, or a prominence with PCTR (with three different values of $\gamma$). With a low $\gamma$, we have a fairly extended transition region, with a lot of prominence material at relatively high temperatures, and therefore where collisional excitations will be important. With a high $\gamma$, we obtain a narrow transition region, and therefore less prominence material at high temperature.
For each of these 4 prominence models, we performed our calculations at different altitudes / radial velocities. Our adopted values for altitude and velocity follow the observations reported in \citet{2007ApJ...663.1354K} in their figure 2 (1998 July 11 event).
The set of plasma parameters used for these computations is therefore as follows:
\begin{itemize}
	\item central temperature: 8000~K,
	\item temperature at the coronal boundary: 10$^5$~K (PCTR models only),
	\item central pressure: 0.1~$\unit{dyn\,cm^{-2}}$,
	\item pressure at the coronal boundary: 0.01~$\unit{dyn\,cm^{-2}}$ (PCTR models only),
	\item total column mass: $2.8 \times 10^{-6} \unit{g\,cm^{-2}}$,
	\item microturbulent velocity: 5~$\unit{km\,s^{-1}}$.
\end{itemize}
These plasma parameters correspond to typical values for quiescent prominences as reported in the Hvar reference atmosphere table \citep{engvoldetal90} and usually adopted in the modelling of quiescent prominences \citep[e.g.][]{ghv}. 
These values may not be strictly valid to study active and eruptive prominences. However our main objective here is to gain a qualitative understanding of the processes at work in the formation of the hydrogen and helium spectra and of the resulting intensities in lines of interest. 
There is no thermodynamic model of eruptive prominence we have heard of. 
In order to compare our computed intensities with observations, we require a better understanding of the range of values that the plasma parameters actually take in active and eruptive prominences. The situation might improve with the recent work of \cite{2008ApJ...673..611K}, but more studies of this type are needed.

\section{Results}

\subsection{\ion{He}{ii} $\lambda$ 304~\AA}

\begin{figure}[t]
\vspace*{2mm}
\begin{center}
\includegraphics[width=8.3cm]{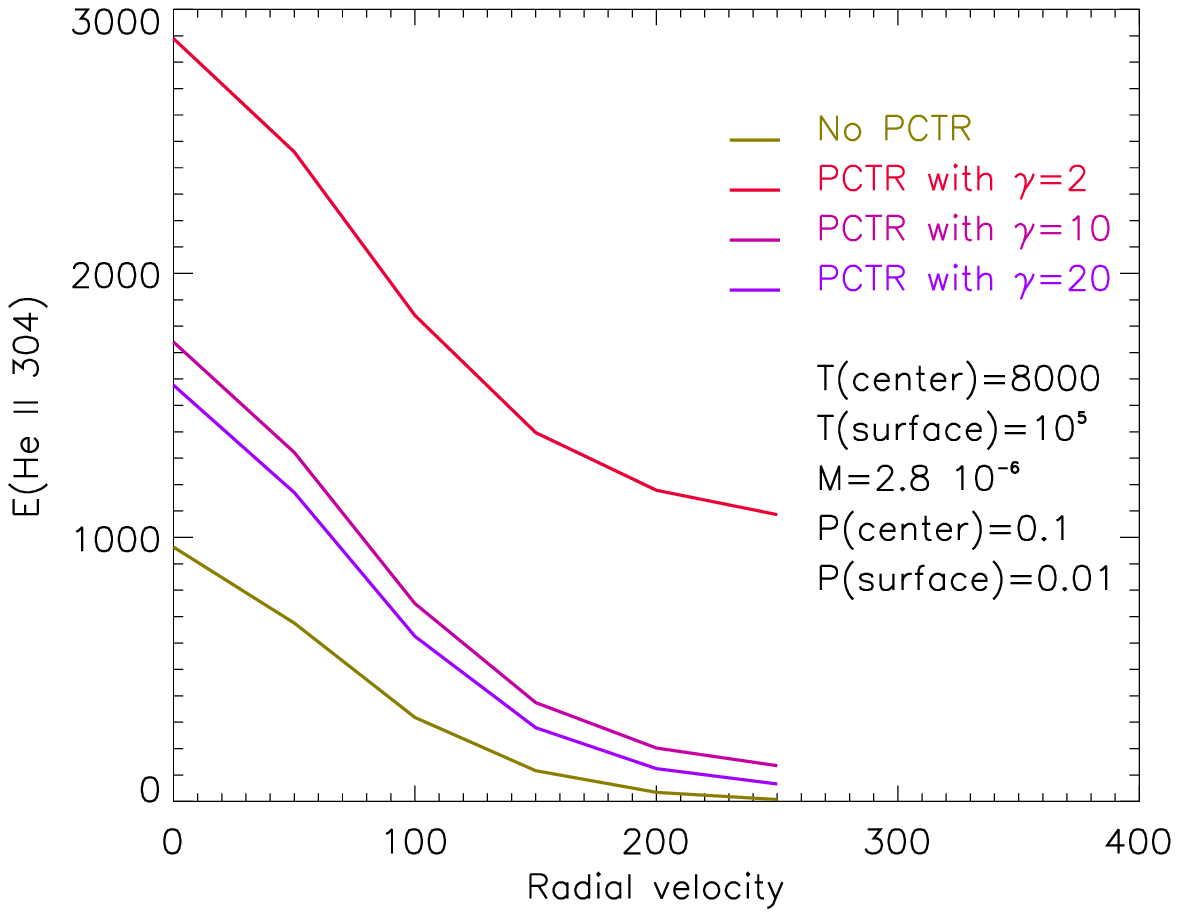}
\end{center}
\caption{Variation of the \ion{He}{ii} $\lambda$ 304~\AA\ line intensity (in $\unit{erg\,s^{-1}\,cm^{-2}\,sr^{-1}\,\AA^{-1}}$) as a function of the radial velocity of the prominence plasma (in $\unit{km\,s^{-1}}$) for 4 different types of prominence atmospheres.}
\label{heii304}
\end{figure}

Figure~2 shows the variation with the radial speed of the integrated intensity of the \ion{He}{ii} 304 line for 4 different prominence conditions. The brown line corresponds to the model without a prominence-to-corona transition region (no PCTR). In this model, the prominence temperature is 8000~K everywhere, and the pressure is also fixed to its value at the centre of the prominence. Under these conditions the \ion{He}{ii} line forms by the scattering of the incident radiation, and this makes it very sensitive to the Doppler dimming effect \citep{2007A&A...463.1171L}. 
The red curve shows the variation of the line intensity for a model with a transition region having a shallow temperature gradient ($\gamma=2$). In that case, the PCTR is fairly extended and the amount of hot material is large. 
It is apparent that the integrated intensity in this line is greatly enhanced, especially at low velocities, due to an important contribution of collisional excitations. 
As the radial velocity increases, the Doppler dimming effect is still visible, although less strong than when the transition region is not included in the calculations. Because of the significant contribution of collisional excitations in this case, the line intensity is still high  at large velocities, even if the radiative excitation is not effective any more as a line formation mechanism.
When  the temperature gradient in the transition region is increased, keeping all other thermodynamical parameters fixed, the transition region becomes less extended, therefore there is less hot material, and the resulting intensity tends towards the isothermal case. 

The most important feature of this figure is that the \ion{He}{ii} resonance line is very sensitive to the Doppler dimming effect in most cases.
Therefore, this study shows that it is crucial to take into account the velocity fields within the prominence if one wants to perform a diagnostic of the plasma using the \ion{He}{ii} resonance lines.
An interesting consequence is that, once a diagnostic is made, the \ion{He}{ii} $\lambda$ 304~\AA\ line can be used to determine the actual direction of propagation of an erupting prominence if this method is used in conjunction with high cadence imaging. 
Indeed, an imager with a good temporal resolution will give us the projection of the velocity upon the plane of the sky along with the direction of propagation, while  our diagnostic method yields both the thermodynamical parameters and the radial velocity.
In addition, the present work confirms the results presented in \citet{2007A&A...463.1171L}, but extending them to the case where a PCTR is included in the prominence model. 
Note that the behaviour of the \ion{He}{ii} $\lambda$ 256~\AA\ resonance line is similar to the 304~\AA\ line.

\begin{figure}[t]
\vspace*{2mm}
\begin{center}
\includegraphics[width=8.3cm]{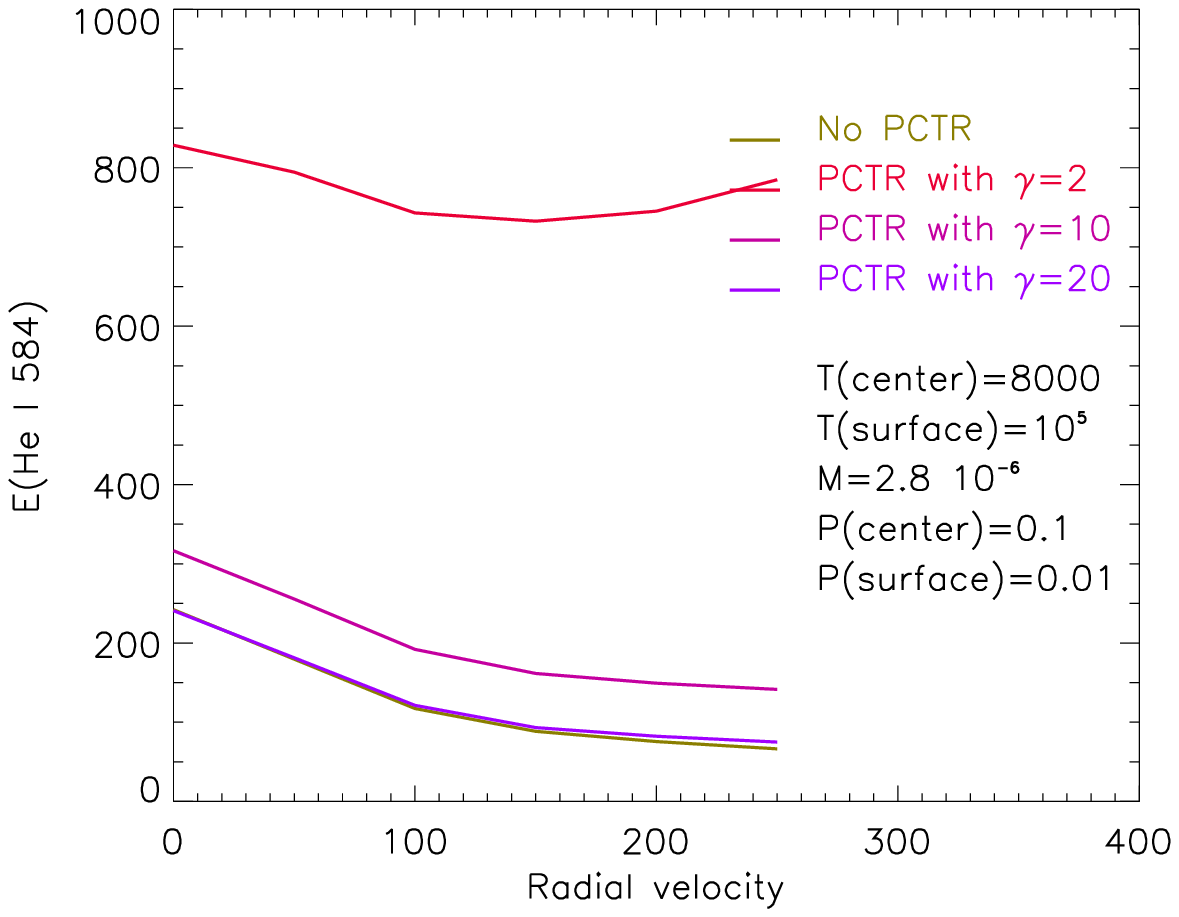}
\end{center}
\caption{Variation of the \ion{He}{i} $\lambda$ 584~\AA\ line intensity (in $\unit{erg\,s^{-1}\,cm^{-2}\,sr^{-1}\,\AA^{-1}}$) as a function of the radial velocity of the prominence plasma (in $\unit{km\,s^{-1}}$) for 4 different types of prominence atmospheres.}
\label{hei584}
\end{figure}

\subsection{\ion{He}{i} $\lambda$ 584~\AA}

	The \ion{He}{i} resonance line at 584~\AA\ has a somewhat more complex behaviour than the \ion{He}{ii} resonance line (Fig.~3). When the transition region is not too extended along the LOS, its intensity mainly decreases with speed, but the contribution of collisional excitations at large velocities becomes more important than the radiative contribution in the line formation, and the intensity is less sensitive to the Doppler effect. When the PCTR is extended, the combination of collisional excitations due to the presence of hot material, and the coupling with other transitions, lead to some Doppler brightening at large velocities.
	This shows that the \ion{He}{i} 584 resonance line is mostly sensitive to Doppler dimming for velocities up to 100 $\unit{km\,s^{-1}}$. Again,  the consideration of velocity fields in the modelling is essential.

	\subsection{H$\alpha$}

	\begin{figure}[t]
	\vspace*{2mm}
	\begin{center}
	\includegraphics[width=8.3cm]{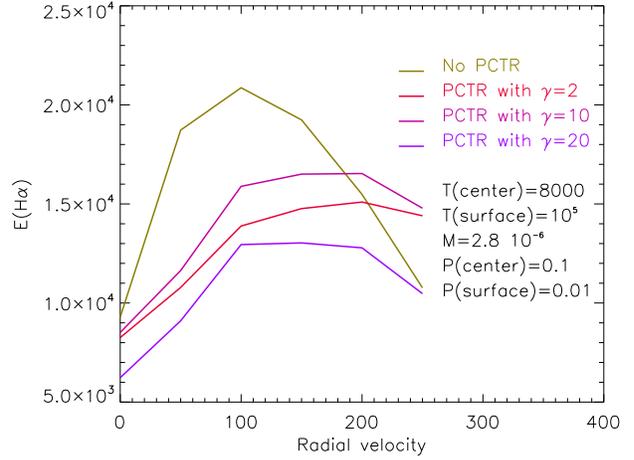}
	\end{center}
	\caption{Variation of the hydrogen H$\alpha$  line intensity (in $\unit{erg\,s^{-1}\,cm^{-2}\,sr^{-1}\,\AA^{-1}}$) as a function of the radial velocity of the prominence plasma (in $\unit{km\,s^{-1}}$) for 4 different types of prominence atmospheres.}
	\label{halpha}
	\end{figure}

	As can be seen on figure~4, this hydrogen line has a more complex behaviour than the helium lines. Because of a coupling between the first two excited levels of hydrogen, we observe a combination of Doppler brightening at low velocities, and then Doppler dimming when the radial velocity of the prominence plasma exceeds 100~$\unit{km\,s^{-1}}$.
	This reflects the fact that the population of the upper level of the H$\alpha$ transition is depleted because of a decrease in the photo-excitation from the H Ly$\beta$ line (due to the Doppler dimming of this line).
	Because of the complexity of the variation of the H$\alpha$ line intensity with radial speed, it should be discarded for diagnostics at large velocities. However it is a useful line for diagnostics at low velocities. 
	The conclusion is the same as in the two previous cases: it is essential to take into account the velocity fields in non-LTE radiative transfer modelling of active and eruptive prominences.

	\conclusions
	In this work we presented the first synthesis of H and He lines in active and eruptive prominences with a transition region between the prominence and the corona.

	We have shown that the \ion{He}{ii} 304~\AA\ line is strongly dependent on the radial velocity.
	It is thus necessary to carefully consider the role of the radial motion of the plasma in the modelling of this line. The coupling existing between the energy states of neutral and ionized helium and the transitions arising between them leads us to the conclusion that velocity fields must be taken into account for the modelling of any helium line.
	We can also conclude that the inclusion of a prominence-to-corona transition region does not modify the previous results obtained in the isothermal case \citep{2007A&A...463.1171L} despite the increased contribution of collisions in the line formation.
	Other lines such as \ion{He}{i} $\lambda$ 584~\AA\ and H$\alpha$ are also sensitive to the Doppler dimming / brightening due to the radial motion of the plasma at velocities up to $\sim$ 100 $\unit{km\,s^{-1}}$. 
	At higher velocities, the absorption profile of the transition gets out of resonance with the incident radiation, and the resulting variation of the emergent intensity with the radial velocity is dependent on several parameters such as the strength of collisional excitations, coupling with other transitions, etc\ldots

	Further modelling is necessary to take into account the fact that the prominence plasma is heated during an eruption. In this situation, Doppler dimming will be affected by an enhanced contribution of collisional excitation.
	This is one of the major challenge in this type of study: it is essential to make a distinction between a change in the thermodynamic state of the plasma (e.g., an increase in temperature) and a change of the radial velocity. Does a decrease in the intensity of a line formed at low temperatures (say below 10000~K) correspond to a Doppler dimming effect due to the radial motion of the plasma, or to the heating of the plasma?
	To answer this, we have to use other lines to perform an independent diagnostic of the prominence plasma. For instance, the line intensity of the \ion{He}{i} $\lambda$10830~\AA\ line is not very sensitive to the radial velocity \citep{2006IAUJD...3E..47L}. Therefore this line can be used to determine the plasma parameters.
	Another mean of distinguishing between variations in intensity due to a change of the plasma state or to the radial velocity is to use the full line profiles. The radial motion of the prominence induces asymmetries in the line profile which cannot be attributed to temperature or pressure effects \citep{1997A&A...325..803G,2007A&A...463.1171L}.
	We will devote another paper to a more detailed study of the line profiles for the same type of prominence models with PCTR that we have used in the present work.

	It may be premature to draw any definite conclusion about the interpretation of the numerous observations of active and eruptive prominences made by instruments such as EIT on SOHO \citep{papier_eit}. There are very few observational studies giving a quantitative interpretation of this line in prominences.  Again, the question of the contribution of the heating of the plasma versus the Doppler dimming effect has yet to be resolved. This is a complex issue, since many properties of the prominence plasma will vary during the eruption. As we  progress towards more sophisticated modelling, we will be able to disentangle this. We have demonstrated that the \ion{He}{ii} 304~\AA\ line is sensitive to Doppler dimming. How much the prominence observations made by EIT are affected by this remains an open question, but the interpretation of these observations cannot ignore this issue.

	In a future work, our modelling investigations will address active and eruptive filaments. As these structures are seen on the disk, the radial motion of the plasma cannot be inferred from imaging measurements, and Doppler shifts cannot be interpreted in a straightforward manner. 
	The synthesis of spectral lines from H and He resulting from our non-LTE radiative transfer codes with velocity fields will be an asset to study the plasma parameters in these dynamic structures.

	\begin{acknowledgements}
	The authors would like to thank the referees for their comments.
	Part of this work was funded by a grant from STFC to the University of Wales Aberystwyth. NL acknowledges financial support from the Gooding Fund of the University of Wales Aberystwyth and the Plasma Physics Group of the Institute of Physics.
	This research has made use of NASA's Astrophysics Data System.
	\end{acknowledgements}

\end{document}